\documentclass[fleqn,usenatbib]{mnras}

\usepackage{newtxtext,newtxmath}

\usepackage[T1]{fontenc}

\DeclareRobustCommand{\VAN}[3]{#2}
\let\VANthebibliography\thebibliography
\def\thebibliography{\DeclareRobustCommand{\VAN}[3]{##3}\VANthebibliography}

\usepackage{graphicx}	
\usepackage{amsmath}	
\usepackage[defaultcolor=red]{changes}

\title[Tidal Tracks and Artificial Disruption]{Tidal Tracks and Artificial Disruption of Cold Dark Matter Halos}

\author[Andrew J. Benson \& Xiaolong Du]{
Andrew J. Benson,$^{1}$\thanks{E-mail: abenson@carnegiescience.edu}
Xiaolong Du$^{1}$
\\
$^{1}$Carnegie Institution for Science, 813 Santa Barbara Street, Pasadena, CA 91101, USA
}

\date{Accepted XXX. Received YYY; in original form ZZZ}

\pubyear{202}

\begin{document}
\label{firstpage}
\pagerange{\pageref{firstpage}--\pageref{lastpage}}
\maketitle

\begin{abstract}
We describe a simple extension to existing models for the tidal heating of dark matter subhalos which takes into account second order terms in the impulse approximation for tidal heating. We show that this revised model can accurately match the tidal tracks along which subhalos evolve as measured in high-resolution N-body simulations. We further demonstrate that, when a constant density core is introduced into a subhalo, this model is able to quantitatively reproduce the evolution and artificial disruption of N-body subhalos arising from finite resolution effects. Combining these results we confirm prior work indicating that artificial disruption in N-body simulations can result in a factor two underestimate of the subhalo mass function in the inner regions of host halos, and a 10--20\% reduction over the entire virial volume.
\end{abstract}




\section{Introduction}

Understanding the evolution and possible destruction of dark matter subhalos as they orbit within the tidal field of their host halo has important consequences for many astrophysical observables, including substructure lensing, the frequency of gaps in stellar streams, and the rate of dark matter annihilation \citep{Carlberg_2012,10.1093/mnras/stw1957,2019PhRvD.100f3505D,2020MNRAS.492L..12G}. Tidal interactions with the host cause subhalos to lose mass due to tidal stripping, but also heat the regions of the subhalos interior to the tidal radius, causing it to expand. This results in a gradual evolution of the density profile of the subhalo in a way that follows a so-called ``tidal track'' which seems to be independent (for density profiles initially following the NFW form) of the details of the subhalo orbit and mass loss and depends only on the total mass lost (as first shown by \citealt{2008ApJ...673..226P}).

The most recent studies, for example \cite{2021MNRAS.505...18E}, confirm this result. Specifically, \cite{2021MNRAS.505...18E} show, using N-body simulations, that as a halo, initially having an NFW density profile \citep{1996ApJ...462..563N}, evolves in a tidal field the quantity $(r_\mathrm{max}/r_\mathrm{max,0},v_\mathrm{max}/v_\mathrm{max,0})$, where $v_\mathrm{max}$ is the maximum velocity in the rotation curve of the subhalo, $r_\mathrm{max}$ is the radius at which that maximum occurs, and subscript ``0'' indicates the corresponding quantities in the initial density profile, follows a curve which is accurately described by the function:
\begin{equation}
 \frac{v_\mathrm{max}}{v_\mathrm{max,0}} = 2^\alpha \left(\frac{r_\mathrm{max}}{r_\mathrm{max,0}}\right)^\beta \left[ 1 + \left(\frac{r_\mathrm{max}}{r_\mathrm{max,0}}\right)^2 \right]^{-\alpha},
 \label{eq:tidalTrackFit}
\end{equation}
with $\alpha = 0.40$ and $\beta=0.65$ independent of the parameters of the subhalo orbit.

Prior semi-analytic work \citep{2001ApJ...559..716T,2014ApJ...792...24P} has modelled the tidal evolution of subhalo density profiles using a simple model of tidal heating. Typically, this has utilized the approach of \cite{1999ApJ...514..109G} who derived the rate of heating using an impulse approximation, plus adiabatic corrections, and validated their results against N-body simulations. \cite{2003MNRAS.341..434T} employed the model of \cite{1999ApJ...514..109G}, including the second-order heating terms, in constructing a semi-analytic model for subhalo evolution, which they demonstrated could accurately reproduce results from N-body experiments. \cite{2019PhRvD.100f3505D} tested the model of \cite{2014ApJ...792...24P} against a wider range of N-body simulations, finding it to be ``remarkably accurate'', although unable to capture the full dependence on orbital and structural parameters found in the simulations. Lastly, \cite{2010ApJ...709.1138D} used analytic estimates of tidal heating to explore the effects of tidal shocking by galactic disks on the survival of subhalos, finding that this process could reduce the abundance of subhalos by large factors in regions close to massive galaxies. These works make it clear that tidal heating is a crucial ingredient in modeling the evolution and destruction of subhalos.

In \S\ref{sec:methods} we will show that this approach, in its usual form, does not accurately match the tidal tracks found by \cite{2021MNRAS.505...18E}, but by including a second order term in the tidal heating rate we are able to accurately reproduce these tidal tracks. In \S\ref{sec:results}, we will show that this revised model is also able to match the results of N-body experiments exploring the tidal destruction of subhalos due to the finite mass and spatial resolution of the simulation. This allows us to examine how such artificial disruption effects have impacted the results obtained from high resolution cosmological simulations of subhalo populations. In \S\ref{sec:discussion} we discuss the implications of these results, and draw our conclusions in \S\ref{sec:conclusions}.

\section{Methods}\label{sec:methods}

To explore the tidal evolution of subhalos using semi-analytic methods we make use of the subhalo orbital physics in the Galacticus model \citep{2012NewA...17..175B}. These orbital physics, including the tidal heating model which is most relevant for this work, are described in detail in \cite{2014ApJ...792...24P}, and are based on the model first proposed by \cite{2001ApJ...559..716T}.

Briefly, it is assumed that each spherical shell of matter in the subhalo receives some heating (i.e. energy added by tidal effects), corresponding to a change in specific energy of $\Delta \epsilon$, and responds by expanding. Under the assumption of no shell crossing the final radius, $r_\mathrm{f}$, of the shell is related to its initial radius, $r_\mathrm{i}$, by
\begin{equation}
\Delta\epsilon = \frac{\mathrm{G}M_\mathrm{i}}{2 r_\mathrm{i}}-\frac{\mathrm{G}M_\mathrm{i}}{2 r_\mathrm{f}},
\end{equation}
where $M_\mathrm{i}$ is the mass contained within the shell.

To predict the resulting density profile we then need to compute $\Delta \epsilon (r_\mathrm{i})$.

\subsection{Original Tidal Heating Model}\label{sec:originalModel}

The tidal heating model described by \cite{2014ApJ...792...24P}, and which largely follows that of \cite{2001ApJ...559..716T}, assumes a heating rate per unit mass of \citep{1999ApJ...514..109G}
\begin{equation}
\Delta\dot{\epsilon}(r) = \frac{\epsilon_\mathrm{h}}{3} \left[1+(\omega_\mathrm{p} T_\mathrm{shock})^2\right]^{-\gamma} r^2 g_{ab} G_{ab}
\label{eq:heatingFirstOrder}
\end{equation}
where $\epsilon_\mathrm{h}$ is a normalization coefficient, $\omega_\mathrm{p}$ is the angular frequency of particles at the half mass radius of the satellite, $T_\mathrm{shock}$ is the tidal shock time-scale, $g_{ab}$ is the tidal tensor, and $G_{ab}$ is the time integral of that tensor. 

\cite{2020MNRAS.498.3902Y} calibrated this model to match the results of the Caterpillar \citep{2016ApJ...818...10G} and ELVIS \citep{2014MNRAS.438.2578G} high-resolution cosmological simulations of subhalos\footnote{Specifically, \protect\cite{2020MNRAS.498.3902Y} calibrated to the $z=0$ subhalo bound mass function, and the $v_\mathrm{max}$ vs. $M_\mathrm{subhalo}$ relation. They did \emph{not} calibrate to the radial distribution of subhalos, nor to the peak mass function from those simulations, both of which can be significantly affected by resolutions effects (\protect\citealt{2021MNRAS.503.4075G}, Nadler et al., in prep.). As such, attempting to calibrate to those functions without a careful treatment of resolutions effects, such as developed in this work, could lead to biased results.}, and found a best-fit value of $\epsilon_\mathrm{h}=5.3^{+1.8}_{-1.6}$ when fitting to the Caterpillar simulations under the assumption of $\gamma=2.5$. The exponent $\gamma$ controls the adiabatic correction term, discussed in detail by \cite{1999ApJ...513..626G}, in the above equation. The value of $\gamma$ is somewhat uncertain---\cite{1999ApJ...513..626G} find a value of $\gamma = 2.5$ (which was used by \citealt{2014ApJ...792...24P}), while theoretical considerations predict $\gamma=1.5$ in the slow-shock regime \citep{1999ApJ...513..626G,1994AJ....108.1398W,1994AJ....108.1403W}. \cite{2020MNRAS.498.3902Y} found that $\gamma = 0$ resulted in a better match to the N-body $v_\mathrm{max}$ vs. $M_\mathrm{subhalo}$ relation, although, as we will discuss below, the calibration of \cite{2020MNRAS.498.3902Y} should be revised in light of the results in this work. In this work we use the value $\gamma=1.5$ appropriate for slow shocks. However, it is important to note that neither $\gamma$ nor $\epsilon_\mathrm{h}$ will affect the form of the tidal track predicted by this model, which depends only upon the radial dependence of the heating $\Delta \epsilon(r) \propto r^2$. These parameters instead affect how quickly a given subhalo will move along its tidal track.

To explore the tidal track resulting from this heating model we set up a subhalo-host halo system in Galacticus which closely matches that used by \cite{2021MNRAS.505...18E}. Specifically, the subhalo is modelled as an NFW profile, exponentially-truncated beyond $10 r_\mathrm{s}$ (see, for example, \citealt{2004ApJ...608..663K}), while the host halo has an isothermal density profile, $\rho(r) \propto r^{-2}$. The subhalo has a mass within its initial $r_\mathrm{max}$ of $10^6 \mathrm{M}_\odot$, and a scale radius of $r_\mathrm{s}=0.22$~kpc, resulting in $r_\mathrm{max}=0.48$~kpc, and $v_\mathrm{max}=3.0$~km/s. The host halo has a virial mass of $3.7 \times 10^{12} \mathrm{M}_\odot$. The subhalo is initially placed at the apocenter of its orbit at 200~kpc from the center of the host, and given an initial velocity of 50~km/s resulting in a pericenter of approximately 25~kpc. The system is then evolved using the orbital physics model described by \cite{2014ApJ...792...24P} and \cite{2020MNRAS.498.3902Y} for 100~Gyr to allow us to probe into the deeply-tidally-stripped regime. At each time the density profile of the subhalo is computed using the model described above, and from this density profile the instantaneous values of $r_\mathrm{max}$ and $v_\mathrm{max}$ are computed.

\begin{figure}
    \centering
    \includegraphics[width=85mm]{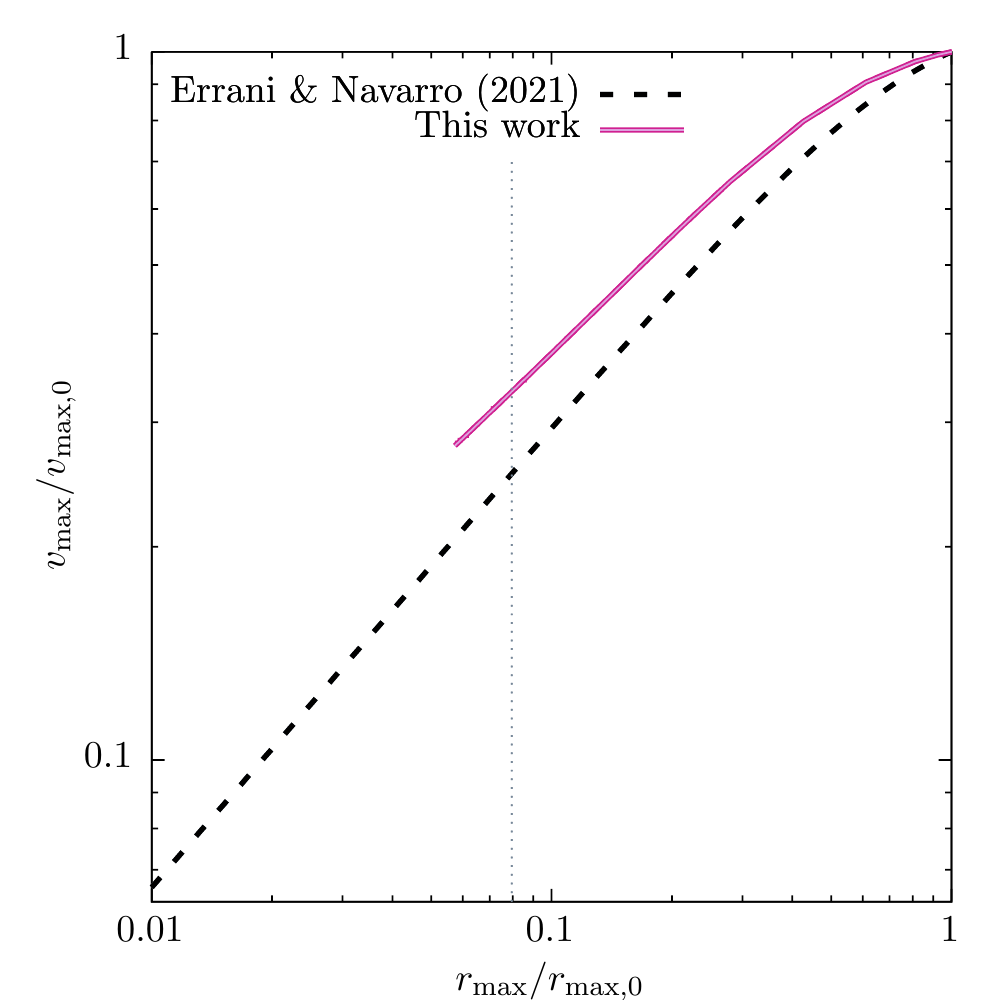}
    \caption{The tidal track of an NFW halo resulting from our first-order tidal heating model is shown by the solid purple line. The dashed black line shows the fitting function of \protect\cite{2021MNRAS.505...18E} which was calibrated to the results of their high-resolution N-body simulations. The vertical dotted line indicates the smallest $r_\mathrm{max}/r_\mathrm{max,0}$ for which \protect\cite{2021MNRAS.505...18E} showed N-body results (in their Figure~6).}
    \label{fig:tidalTrackFirstOrder}
\end{figure}

The results of this experiment are shown in Figure~\ref{fig:tidalTrackFirstOrder}. The solid purple line shows the tidal track resulting from our model, while the black dashed line is the fitting function of \cite{2021MNRAS.505...18E} as defined in equation~(\ref{eq:tidalTrackFit}). The vertical dotted line indicates the smallest $r_\mathrm{max}/r_\mathrm{max,0}$ for which \protect\cite{2021MNRAS.505...18E} showed N-body results (in their Figure~6).

Clearly the tidal heating model described above does not result in a tidal track which matches that found in the high-resolution N-body simulations of \cite{2021MNRAS.505...18E}, it begins to deviate immediately as the subhalo evolves from the upper right corner of the $(r_\mathrm{max}/r_\mathrm{max,0},v_\mathrm{max}/v_\mathrm{max,0})$ plane and has a shallower slope than equation~(\ref{eq:tidalTrackFit}) such that the mismatch grows as the subhalo becomes more and more tidally heated. We reiterate that this result can not be changed by altering the values of either $\gamma$ or $\epsilon_\mathrm{h}$ in this model, which determine only how quickly a subhalo moves along this track as it evolves in a tidal field.

\subsection{An Improved Tidal Heating Model}

The heating rate given by equation~(\ref{eq:heatingFirstOrder}) is derived from the work of \cite{1999ApJ...514..109G}, but accounts for only the first-order perturbation that they derived to the energies of subhalo particles as they experience an impulsive tidal shock. However, as shown by \cite{1995ApJ...438..702K}, the second-order perturbation is of comparable magnitude to the first-order perturbation\footnote{Higher order terms can be of comparable magnitude also, but we ignore them here.}. As such, this second-order term can not be ignored, even in the limit of weak heating.

\citeauthor{2014ApJ...792...24P}~(\citeyear{2014ApJ...792...24P}; see also \citealt{2001ApJ...559..716T}) assume that the second-order term can be accounted for by effectively rolling it in to the normalization parameter, $\epsilon_\mathrm{h}$. While this seems to work reasonably well for the overall evolution (i.e. total mass loss), as shown by \cite{2020MNRAS.498.3902Y}, it does not produce the correct tidal track as shown in the previous section, as the radial dependence of the second-order term does not follow the same $r^2$ behaviour as that of the first-order term.

We therefore attempt to improve this model by explicitly accounting for the second-order perturbation, $\langle E^2 \rangle$, to subhalo particle energies. Using the results from \cite{1999ApJ...514..109G} we now write $\epsilon_2(r) = f_2 \langle E^2 \rangle^{1/2}$, and then write the total perturbation as:
\begin{equation}
\Delta\epsilon(r) = \Delta \epsilon_1 (r) + \Delta \epsilon_2 (r) = \Delta \epsilon_1 (r) + \sqrt{2} f_2 (1+\chi_\mathrm{v}) \sqrt{\Delta \epsilon_1(r) \sigma^2_\mathrm{r}(r)}
\end{equation}
where $\Delta \epsilon_1(r)$ is the first-order term from the original model, $\chi_\mathrm{v}$ is the position-velocity correlation coefficient introduced by \cite{1999ApJ...514..109G}, $\sigma_\mathrm{r}(r)$ is the radial velocity dispersion in the subhalo (prior to any tidal heating), and $f_2$ is a new coefficient introduced to allow us to calibrate the strength of the second-order term. \cite{1999ApJ...514..109G} find that $\chi_\mathrm{v}$ depends weakly on the density profile---we fix it at $\chi_\mathrm{v}=-0.333$ (typical of the values that they found). The precise choice for $\chi_\mathrm{v}$ does not matter as its effect is degenerate with that of $f_2$.

We note that simply adding the second-order term to the first-order term is not justified \emph{a priori}. While the first-order term represents the mean change in energy of the particles, the second-order term describes the mean change in the energy squared, and contributes no net change to particle energies.

Considering this in terms of the distribution function of particle energies, $f(E)$, the second order term effectively acts as a 
smoothing on $f(E)$. Typical cosmological halos have distribution functions that are peak toward large negative (i.e. tightly-bound) energies \citep[e.g.][Figure~5]{2001MNRAS.321..155L}, cutting off toward $E=0$ 
(corresponding to unbound particles).

Smoothing will generally act to reduce the peak at large negative energies, flattening the distribution, reducing the number of tightly-bound particles (and making some particle unbound). As such, we expect the second-order term to result in some expansion of the shell, even though there is no net change in energy. 

Modeling the distribution function in detail, and then solving for a new equilibrium density profile, would be too slow for incorporation of our model into semi-analytic models of subhalo populations. Therefore, we simply add the second-order term to the first-order term. The inclusion of the $f_2$ free parameter allows some calibration of the strength of the second-order term when (mis-)used in this way. This approach will be justified by its success in matching simulation results in the remainder of this paper.

\begin{figure}
    \centering
    \includegraphics[width=85mm]{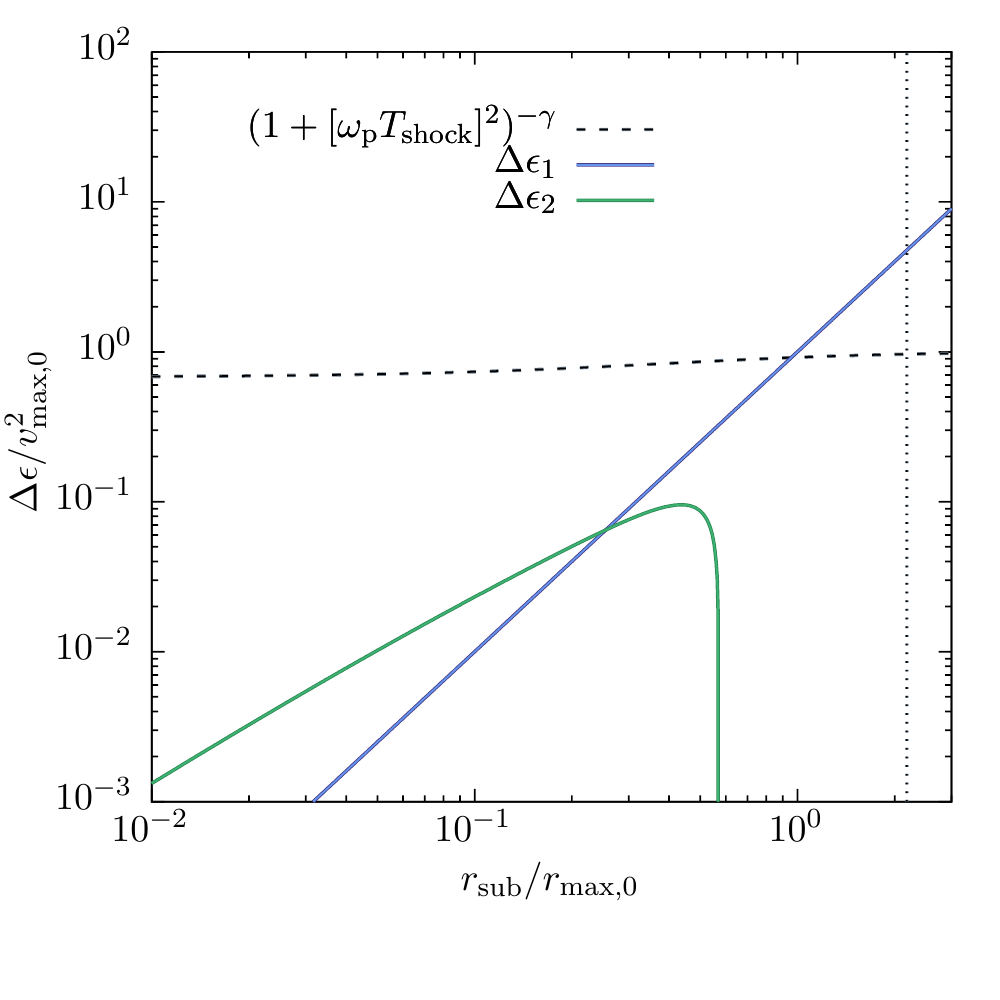}
    \caption{The radial profile of tidal heating in our example subhalo, shown in scaled units of $r_\mathrm{sub}/r_\mathrm{max,0}$ and $\Delta \epsilon/v_\mathrm{max,0}^2$. The blue line shows the first order contribution, $\Delta \epsilon_1$, normalized such that $\Delta \epsilon_1/v_\mathrm{max,0}^2 = 1$ at $r_\mathrm{max}/r_\mathrm{max,0}$, while the green line shows $\Delta \epsilon_2$. The dashed black line indicates the adiabatic correction term, for $T_\mathrm{shock}=0.1$~Gyr, evaluated as a function of radius.}
    \label{fig:radialHeatingProfile}
\end{figure}

The factor of $\sqrt{\Delta\epsilon_1(r) \sigma_\mathrm{r}(r)}$ in the second-order term gives it a different radial dependence than the first-order term. This changes the radial heating profile---generally boosting the heating at smaller radii relative to the original model. We note that the adiabatic correction term in equation~(\protect\ref{eq:heatingFirstOrder}) should also have a radial dependence, since $\omega_\mathrm{p} \sim \sigma_r/r$. We ignore this radial dependence here (since $T_\mathrm{shock}$ is a function of time along the orbit, and so the adiabatic correction term cannot be moved outside of the integral when integrating equation~\ref{eq:heatingFirstOrder}), instead evaluating it at the satellite half-mass radius. In Figure~\ref{fig:radialHeatingProfile} we show the radial profile of heating for our subhalo, with the first- and second-order terms shown by blue and green lines respectively. In this illustrative example we have chosen a normalization such that $\Delta \epsilon_1/v_\mathrm{max,0}^2 = 1$ at $r_\mathrm{max}/r_\mathrm{max,0}$. It can be seen that the second-order term dominates at small radii, but eventually falls below the first-order term (and then truncates rapidly due to the falling velocity dispersion in the outer parts of the halo). Also shown is the adiabatic correction factor (black dashed line), evaluated as a function of radius. It can be seen that this correction factor is only a weak function of radius\footnote{As shown by \protect\cite{2022MNRAS.511.6001E}, the tidal stripping process preferentially removes particles with long orbital periods, and comes to an end when the remaining bound particles in the subhalo have crossing times smaller than a fraction of the orbital timescale at pericentre. Consequently, in the outer parts of the bound remnant the adiabatic correction term will always be of order unity.}. The vertical dotted line indicates the half-mass radius of the halo, at which we evaluate the adiabatic correction term in our model.

\begin{figure}
    \centering
    \includegraphics[width=85mm]{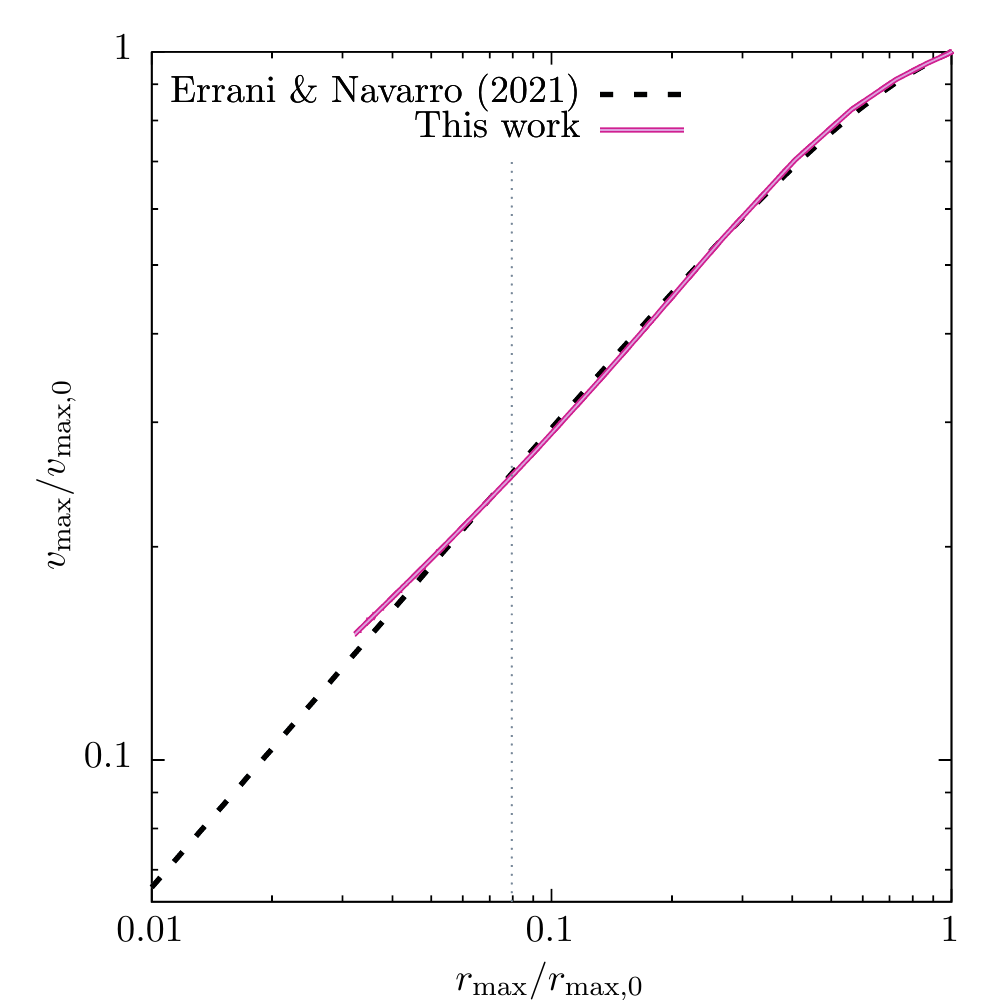}
    \caption{The tidal track of an NFW halo resulting from our second-order tidal heating model is shown by the purple line. The dashed black line shows the fitting function of \protect\cite{2021MNRAS.505...18E} which was calibrated to the results of their high-resolution N-body simulations. The vertical dotted line indicates the smallest $r_\mathrm{max}/r_\mathrm{max,0}$ for which \protect\cite{2021MNRAS.505...18E} showed N-body results (in their Figure~6).}
    \label{fig:tidalTrackSecondOrder}
\end{figure}

We calibrate the value of $f_2$ by simulating the evolution of subhalos orbital evolution for the system described in the previous section (i.e. matched to the system simulated by \citealt{2021MNRAS.505...18E}) for different ratios of pericentric to apocentric radius ($r_\mathrm{p}/r_\mathrm{a}=0.05, 0.10, 0.20$) as considered by \cite{2021MNRAS.505...18E} and seeking the value of $f_2$ which best matches the tidal track given by equation~(\ref{eq:tidalTrackFit}). Optimal matching is obtained for $f_2=0.406$, resulting in the tidal track shown in Figure~\ref{fig:tidalTrackSecondOrder}. The fact that $f_2$ is of order unity is consistent with the expectations for the magnitude of the second-order term based on the analytic arguments of \cite{1995ApJ...438..702K}. With this second-order term included the tidal track predicted by our tidal heating model now accurately matches that found in high-resolution N-body simulations. In particular it reproduces the curvature of the track in the $(r_\mathrm{max}/r_\mathrm{max,0},v_\mathrm{max}/v_\mathrm{max,0})\approx(1,1)$ region, and the transition to a power-law at $(r_\mathrm{max}/r_\mathrm{max,0},v_\mathrm{max}/v_\mathrm{max,0}) < (1,1)$. In the lower-left region of the $(r_\mathrm{max}/r_\mathrm{max,0},v_\mathrm{max}/v_\mathrm{max,0})$ plane our model deviates slightly from equation~(\ref{eq:tidalTrackFit}), but we note that \cite{2021MNRAS.505...18E} did not show N-body results in this region, so equation~(\ref{eq:tidalTrackFit}) is an extrapolation in this range.

\begin{figure}
    \centering
    \includegraphics[width=85mm]{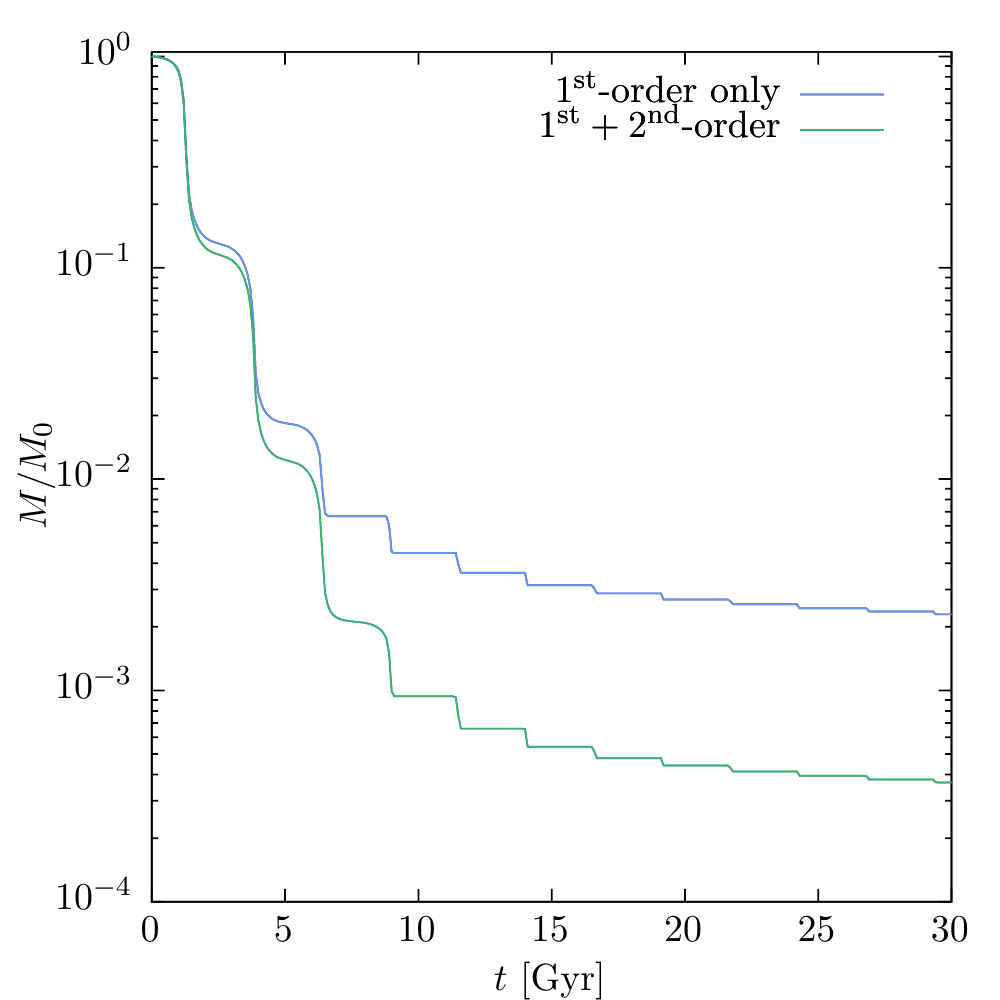}
    \caption{The bound mass, $M$, of the subhalo, normalized to the initial value, $M_0$, as a function of time. The blue line shows results from the original tidal heating model (i.e. including the first-order heating term only), while the green line shows results from our improved model, including both first- and second-order terms.}
    \label{fig:boundMass}
\end{figure}

Figure~\ref{fig:boundMass} shows the bound mass of the subhalo as a function of time for both the original (blue line) and improved (green line) tidal heating models, calculated using the model described in \cite{2014ApJ...792...24P} and \cite{2020MNRAS.498.3902Y}. The additional tidal heating contributed by the second-order term results in greater expansion of the subhalo density profile, which in turn results in more rapid mass loss due to tidal stripping. We caution that the tidal mass loss model used here was calibrated by \cite{2020MNRAS.498.3902Y} to match the results of cosmological N-body simulations using the original tidal heating model. As such, it likely overpredicts the rate of mass loss once the second-order heating term is included. A recalibration of the tidal mass loss model using our improved tidal heating calculation will be presented in a subsequent paper.

\begin{figure}
    \centering
    \includegraphics[width=85mm]{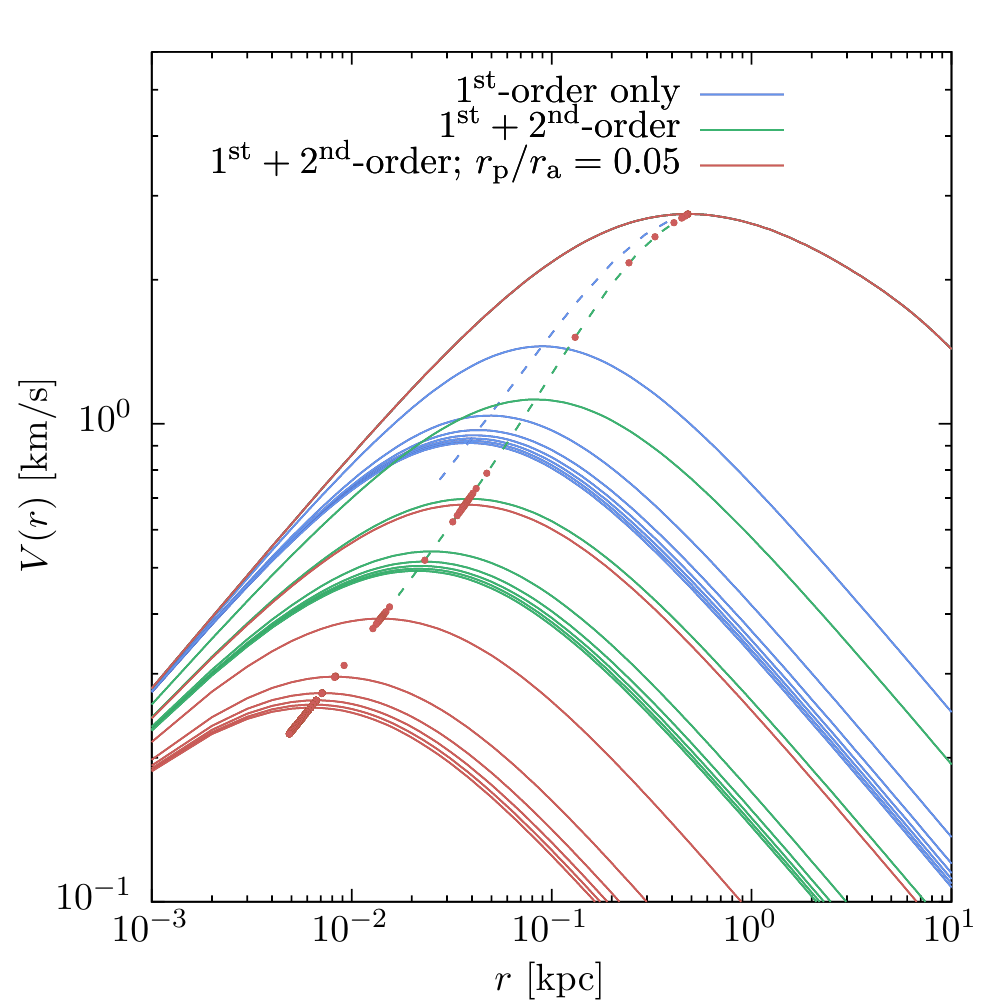}
    \caption{Rotation curves of subhalos experiencing tidal heating. Solid lines show the rotation curves from our original heating model (blue), our improved model (green), and our improved model applied to a subhalo on a more radial orbit (peri- to apo-centric distance ratio of $r_\mathrm{p}/r_\mathrm{a}=0.05$). Rotation curves are shown at times corresponding to each apocenter of the subhalo's orbit. The blue and green dashed lines show the corresponding tidal tracks $(r_\mathrm{max},v_\mathrm{max})$ for the original and improved models. For the more radial orbit, we show the tidal track as red points.}
    \label{fig:rotationCurve}
\end{figure}

As noted in \S\ref{sec:originalModel}, the tidal track predicted by our model is independent of the subhalo orbit\footnote{The orbit \emph{does} determine how quickly the subhalo moves along the tidal track, since, for example, more radial orbits will experience stronger tidal tensors, and so the $G_{ab}$ term in equation~(\ref{eq:heatingFirstOrder}) will grow more rapidly.}---it depends only upon the integrated heating via the $g_{ab} G_{ab}$ term in equation~(\ref{eq:heatingFirstOrder}). Figure~\ref{fig:rotationCurve} shows the rotation curves of subhalos as they are tidally heated---we plot rotation curves at times corresponding to successive apocenters of the subhalo's orbit. The solid lines show the rotation curves from our original heating model (blue), our improved model (green), and our improved model applied to a subhalo on a more radial orbit (peri- to apo-centric distance ratio of $r_\mathrm{p}/r_\mathrm{a}=0.05$). The blue and green dashed lines show the corresponding tidal tracks $(r_\mathrm{max},v_\mathrm{max})$ for the original and improved models. For the more radial orbit, we show the tidal track as red points. It can be seen that, while the effects of tidal heating are more extreme for the more radial orbit (i.e. its rotation curve reduced much further) it nevertheless follows the same tidal track as the less radial orbit.

\section{Results}\label{sec:results}

Having demonstrated that, by including the second-order term in our tidal heating model, we can accurately match the tidal tracks measured from  N-body simulations of NFW subhalos, we now explore how this model can be used to examine finite resolution effects in such simulations.

\citeauthor{{2020MNRAS.491.4591E}}~(\citeyear{{2020MNRAS.491.4591E}}; see also \citealt{2018MNRAS.474.3043V} and \citealt{2018MNRAS.475.4066V}) demonstrate that, in the limit of infinite resolution, NFW density profiles would never be completely destroyed by tidal effects---a tiny subhalo would always remain due to the central density cusp. In practice, any N-body simulation has a finite resolution---both in terms of mass (the particle mass), and length (a softening length or grid scale)---which results in artificial disruption of subhalos as they approach the limits of this resolution. Such artificial disruption could bias measures of the statistical properties of subhalos measured from N-body simulations \citep{2021MNRAS.503.4075G,2021MNRAS.505...18E}.

In \S\ref{sec:finiteResolutionIdealized} we explore how our model can capture these finite resolution effects and the resulting artificial disruption in idealized simulations. We then apply this, in \S\ref{sec:finiteResolutionCosmological}, to study how artificial disruption may be impacting subhalo mass function measurements made from high-resolution cosmological N-body simulations.

\subsection{Finite Resolution Effects in Idealized simulations}\label{sec:finiteResolutionIdealized}

With our model for tidal heating validated and calibrated we now examine the effects of a finite resolution (in length and/or mass) in N-body simulations on the evolution of the tidal track. \cite{2021MNRAS.505...18E} perform simulations to explore the effects of resolution in which they vary the properties of a single subhalo orbiting in a host potential.

We recreate the initial conditions of \cite{2021MNRAS.505...18E} in Galacticus, and introduce a finite radius core into the dark matter profile of the subhalo to mimic the effects of the finite resolution of their N-body simulations. Specifically, the density profile inside the truncation radius is given by:
\begin{equation}
 \rho(r) = \rho_\mathrm{NFW}(r) \left( 1 + \left[ \frac{\Delta x}{r} \right]^2 \right)^{-1/2},
 \label{eq:densityProfileFiniteResolution}
\end{equation}
where $\Delta x$ is a measure of the resolution in the simulation, which we set to be the larger of $\epsilon \Delta x_\mathrm{grid}$ and $r(N m_\mathrm{p})$, where $\Delta x_\mathrm{grid}$ is the size of a grid cell\footnote{In the highest resolution simulations carried out by \protect\cite{2021MNRAS.505...18E} the size of the grid cells is $r_\mathrm{max,0}/128$, which corresponds to around $0.018$ times the mean interparticle spacing for an equivalent cosmological simulation. For a non-grid based calculation of gravitational accelerations an appropriate scale would be the gravitational softening length.} used in the calculation of gravitational forces in \cite{2021MNRAS.505...18E}, $m_\mathrm{p}$ is the particle mass in the N-body simulation, $r(M)$ is the radius in the initial subhalo density profile enclosing a mass $M$, and $\epsilon$ and $N$ are dimensionless parameters (controlling the impact of finite spatial, and finite mass resolution respectively) that we will calibrate to match the results of \cite{2021MNRAS.505...18E}. Note that the density profile described by equation~(\ref{eq:densityProfileFiniteResolution}) has a constant density core on scales below\footnote{More precisely, the density becomes constant for radii, $r \ll \hbox{min}(\Delta x, r_\mathrm{s})$. For the models comparing with the results of \cite{2021MNRAS.505...18E} considered in this section $\Delta x$ is always much smaller than $r_\mathrm{s}$ (even in the least well-resolved model $\Delta x / r_\mathrm{s} \approx 0.03$), such that the central cusp is always well-resolved in the initial conditions.} $\Delta x$. The velocity dispersion in the halo is computed assuming Jeans equilibrium and an isotropic velocity distribution.

We note that this approach of initializing the subhalo with a core, differs from that of \cite{2021MNRAS.505...18E} where the subhalo is initialized with an NFW density profile. The assumption made in this work is that a core will rapidly develop once an N-body simulation of an NFW subhalo begins, making our choice to initialize with such a core reasonable. \cite{2020MNRAS.491.4591E} show, in a controlled N-body experiment, that a core is already well-established in an initially cuspy subhalo after a time corresponding to 4 orbital timescales at the subhalo scale radius (approximately 0.2~Gyr). This is much shorter than the duration of our simulations, justifying the approximation of initializing the subhalos with a core.

\begin{figure*}
\begin{tabular}{cc}
\includegraphics[width=82mm,trim={0 1cm 0 0},clip]{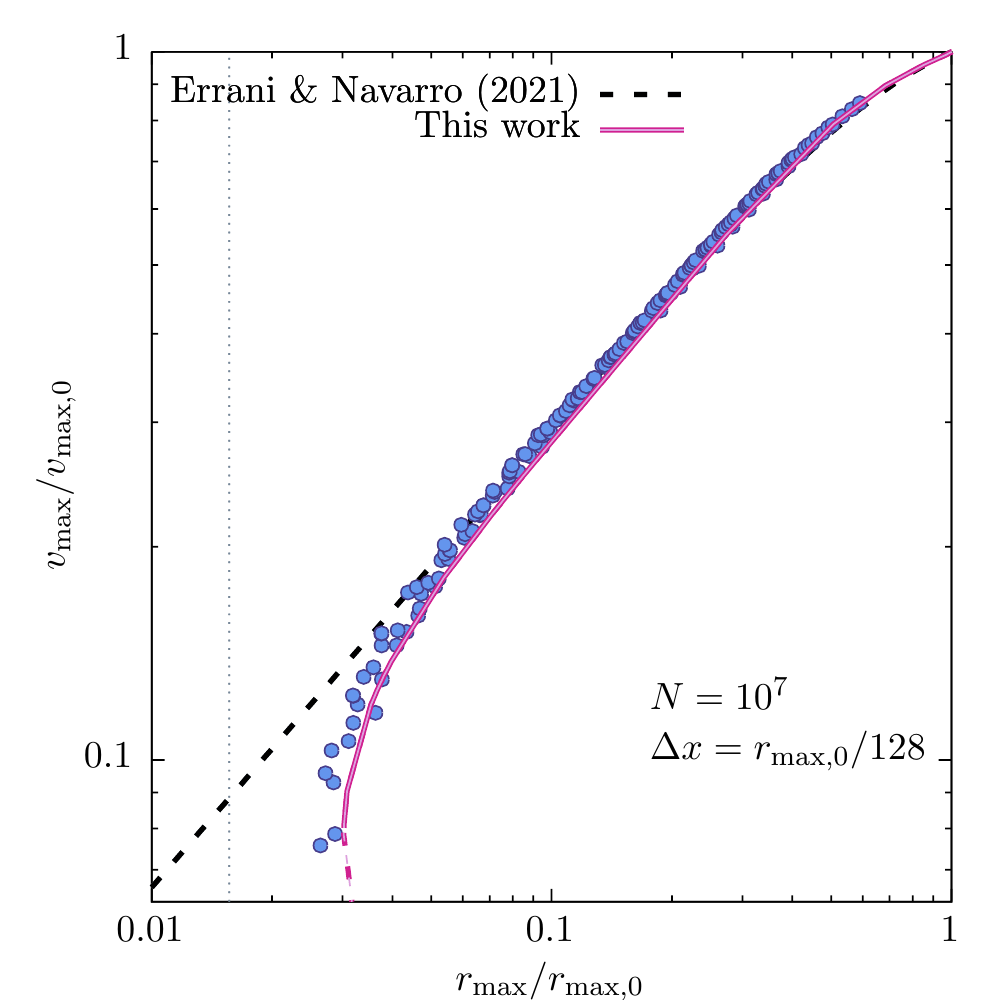} & \includegraphics[width=82mm,trim={0 1cm 0 0},clip]{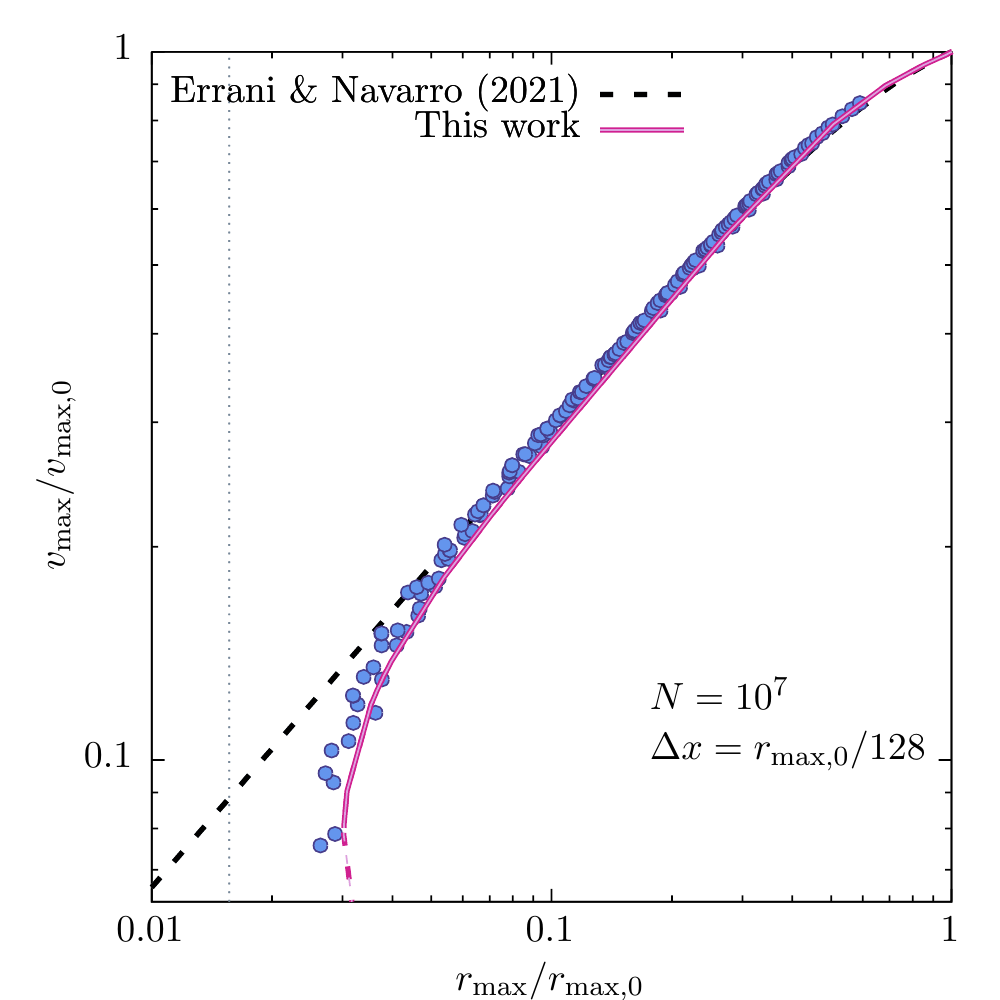}\vspace{-5.5mm} \\
\includegraphics[width=82mm,trim={0 1cm 0 0},clip]{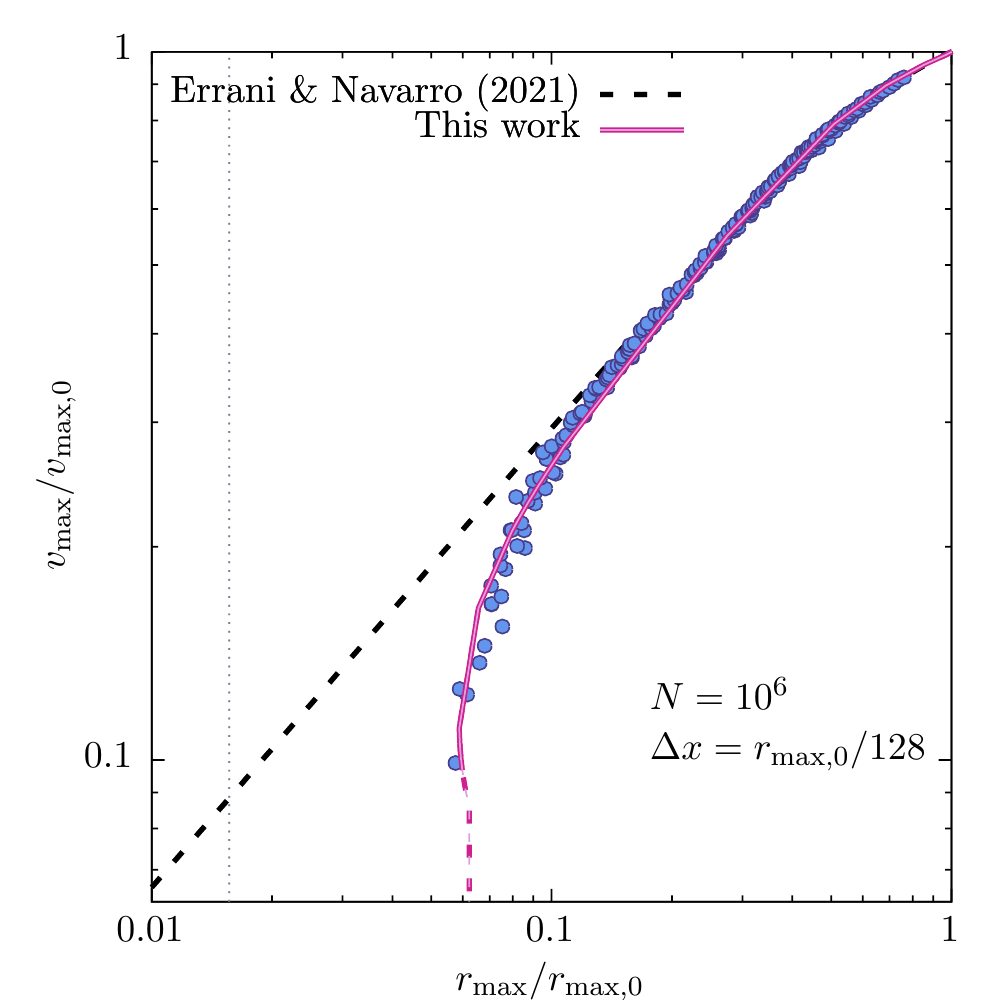} & \includegraphics[width=82mm,trim={0 1cm 0 0},clip]{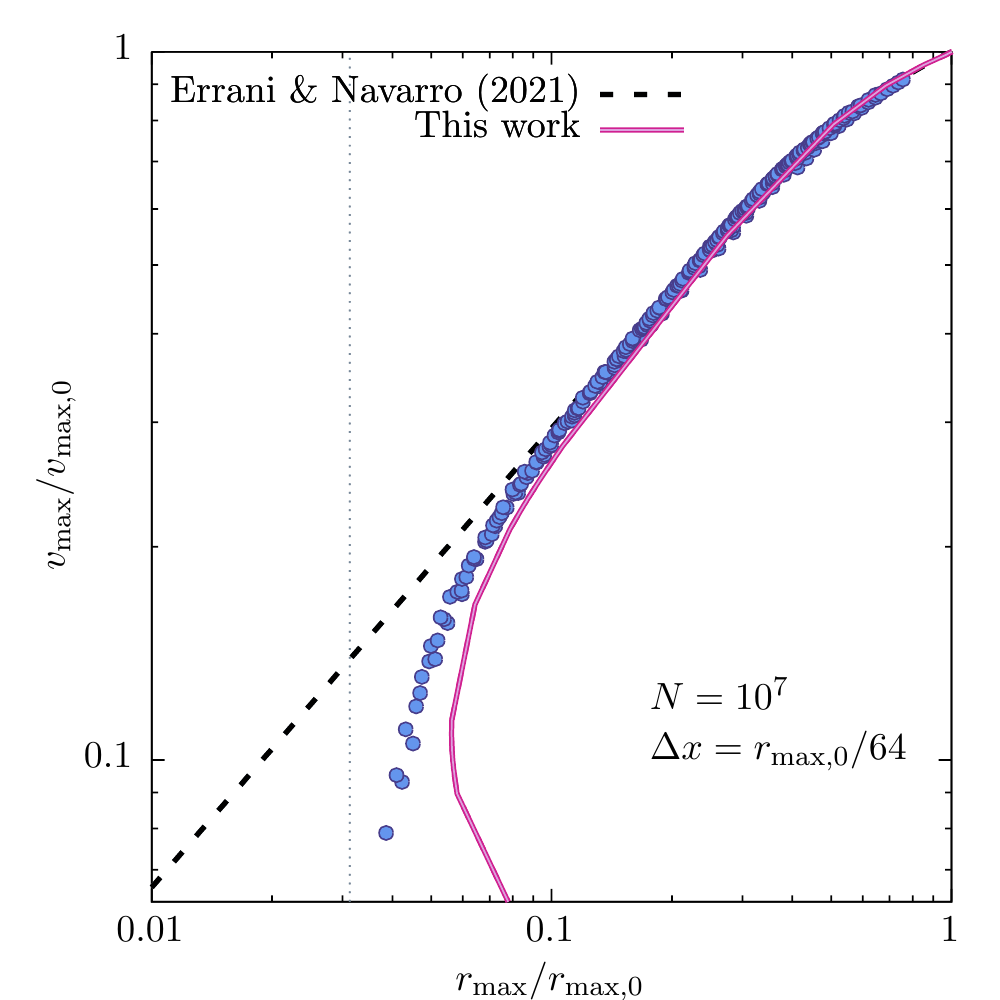}\vspace{-5.5mm} \\
\includegraphics[width=82mm]{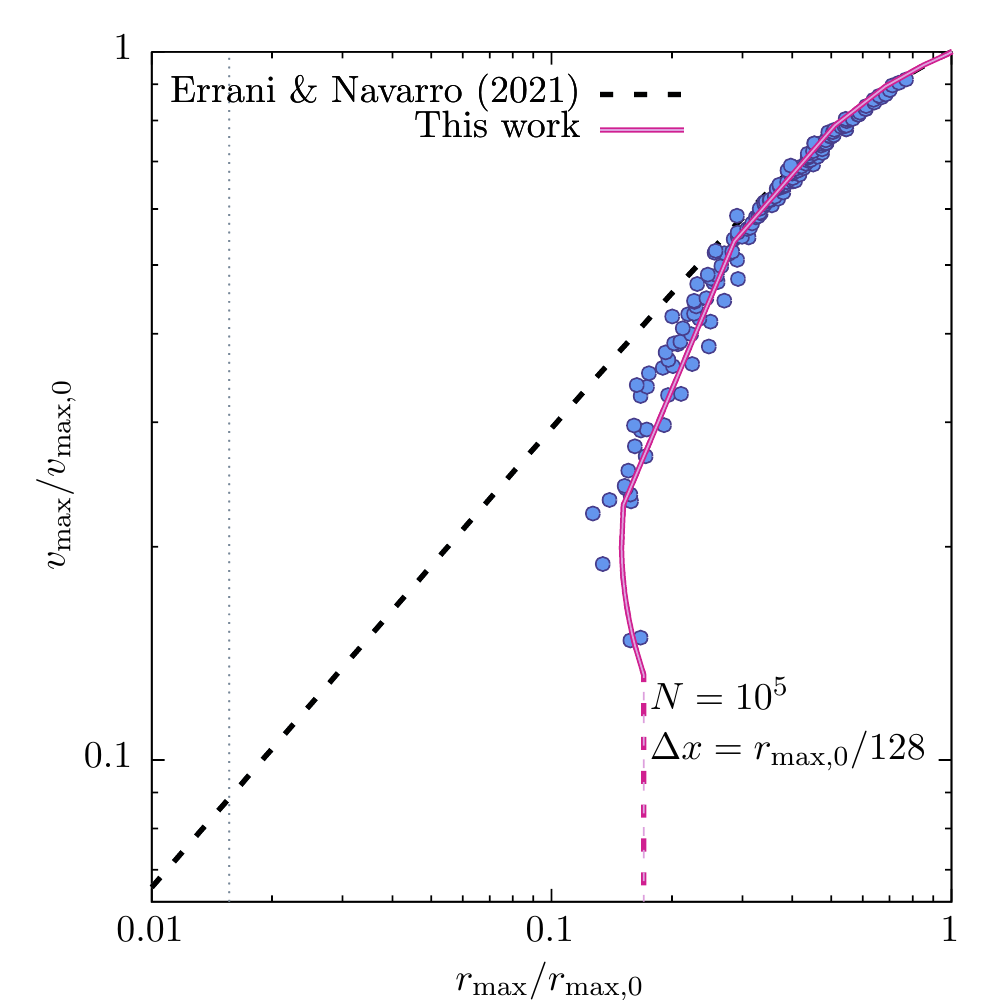} & \includegraphics[width=82mm]{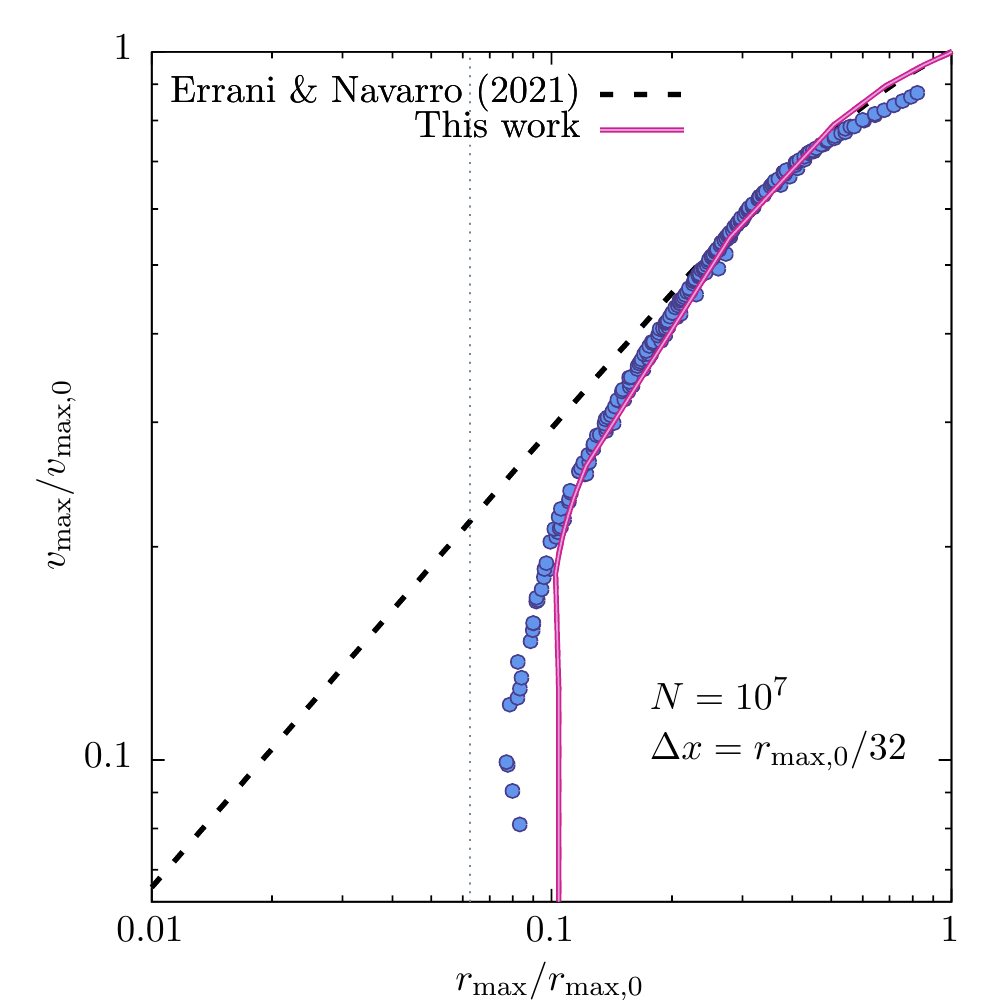} \\
\end{tabular}
\caption{Tidal tracks for subhalos simulated with different mass (left column) and spatial (right column) resolutions. In each panel the dashed line shows the fitting function of \protect\cite{2021MNRAS.505...18E} for a perfectly-resolved subhalo, while blue points show results from \protect\cite{2021MNRAS.505...18E} for subhalos resolved with $N$ particles, and a spatial grid with cell size $\Delta x$ as given in each panel. The purple line shows results from this work using the second order tidal heating algorithm and cored density profiles tuned to match the N-body results.}
\label{fig:tidalTrackResolution}
\end{figure*}

We find that $\epsilon=0.18$ and $N=36$ result in accurate matches to the results of \cite{2021MNRAS.505...18E}. Figure~\ref{fig:tidalTrackResolution} shows tidal tracks for six models matched to simulations of varying mass and spatial resolution performed by \cite{2021MNRAS.505...18E}. The total number of particles in the subhalo and the length resolution are shown in each plot title. The dashed black line is the fit to the tidal track for well-resolved halos from \cite{2021MNRAS.505...18E} as given in equation~(\ref{eq:tidalTrackFit}). The vertical dotted line shows $2\Delta x_\mathrm{grid}$. The blue points are the N-body results from \cite{2021MNRAS.505...18E}, and the red line is from our model including the finite-resolution core. This line is solid for regions where the bound mass of the subhalo is greater than $3,000 m_\mathrm{p}$ and is dashed for regions where the bound mass is less than this.

In all cases our tidal heating model, applied to subhalos with a constant density core matched to the resolution of the corresponding N-body simulation, accurately matches the deviation from the infinite-resolution tidal track that occur as the finite resolution of the simulation begins to affect the subhalo evolution.

\subsection{Finite Resolution Effects in the Caterpillar Simulations}\label{sec:finiteResolutionCosmological}

Using our model for tidal tracks, and the calibrated treatment of finite resolution effects in N-body simulations described in the prior section, we have run realizations of the evolution of the subhalo population in a cosmological halo using the approach described by \cite{2014ApJ...792...24P} and \cite{2020MNRAS.498.3902Y}. We match the cosmological parameters of the Caterpillar LX14 suite of halos \citep{2016ApJ...818...10G} using Galacticus, simulating a total of 17,920 merger trees matched in mass to the $z=0$ halos in the Caterpillar LX14 suite. In these calculations we resolved subhalos down to a mass of $M_\mathrm{res} = 2.9845 \times 10^6\mathrm{M}_\odot$ (100 times the Caterpillar LX14 particle mass). We perform this calculation twice---once with infinite resolution (i.e. pure NFW halos for the subhalos prior to tidal heating), and once including the finite resolution core as described above utilizing the Caterpillar LX14 particle mass ($m_\mathrm{p}=2.9854 \times 10^4 \mathrm{M}_\odot$) and softening length ($r_\mathrm{soft}=76 h^{-1}$~pc comoving) and the best-fit calibration described above, that is we use $\Delta x = \hbox{min}[\epsilon r_\mathrm{soft}, r(N m_\mathrm{p})]$ with the values of $\epsilon$ and $N$ calibrated to the \cite{2021MNRAS.505...18E} simulations\footnote{For the lowest mass halos this may result in $\Delta x \sim r_\mathrm{s}$ or larger. As such, these lowest mass subhalos will not be well-resolved. This is precisely the limitation of cosmological simulations that we are attempting to mimic by introducing an artificial core.}. From these simulations we measure the radial distribution of subhalos in several intervals of bound mass.

\begin{figure*}
\begin{tabular}{cc}
\includegraphics[width=85mm]{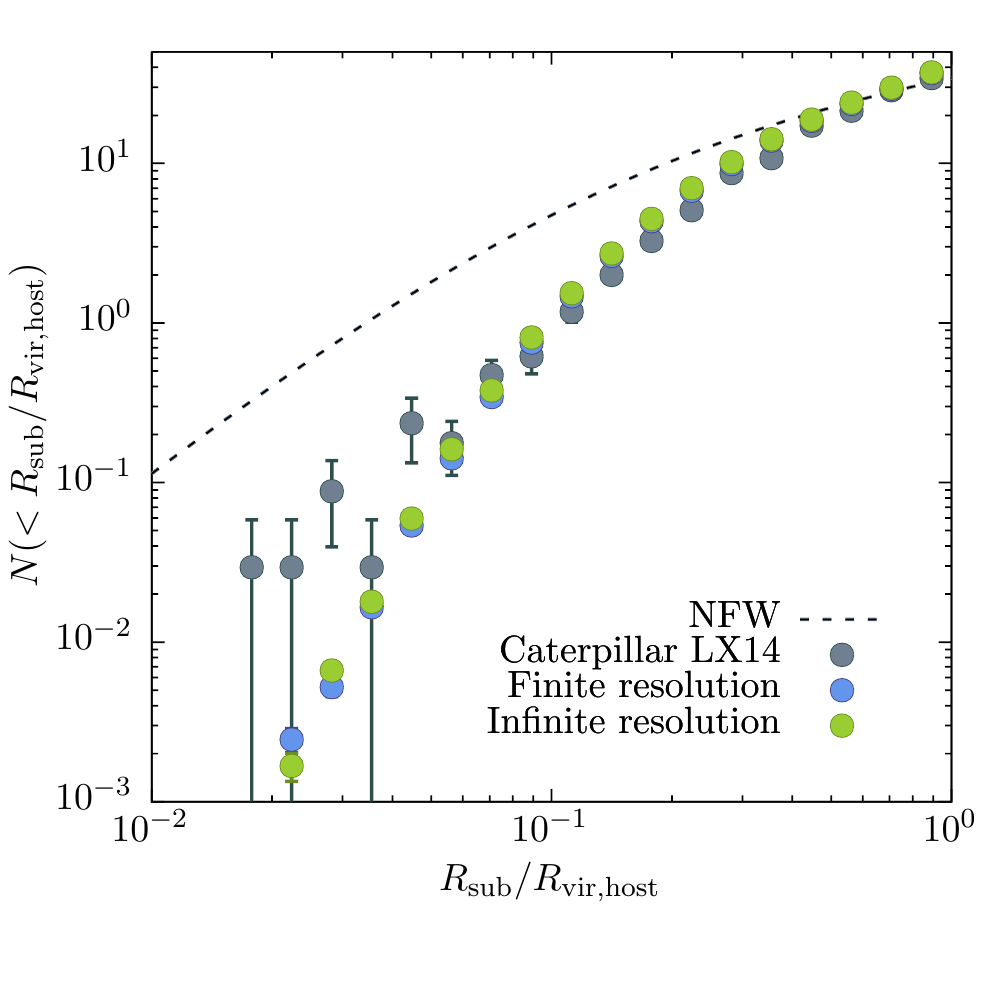} & \includegraphics[width=85mm]{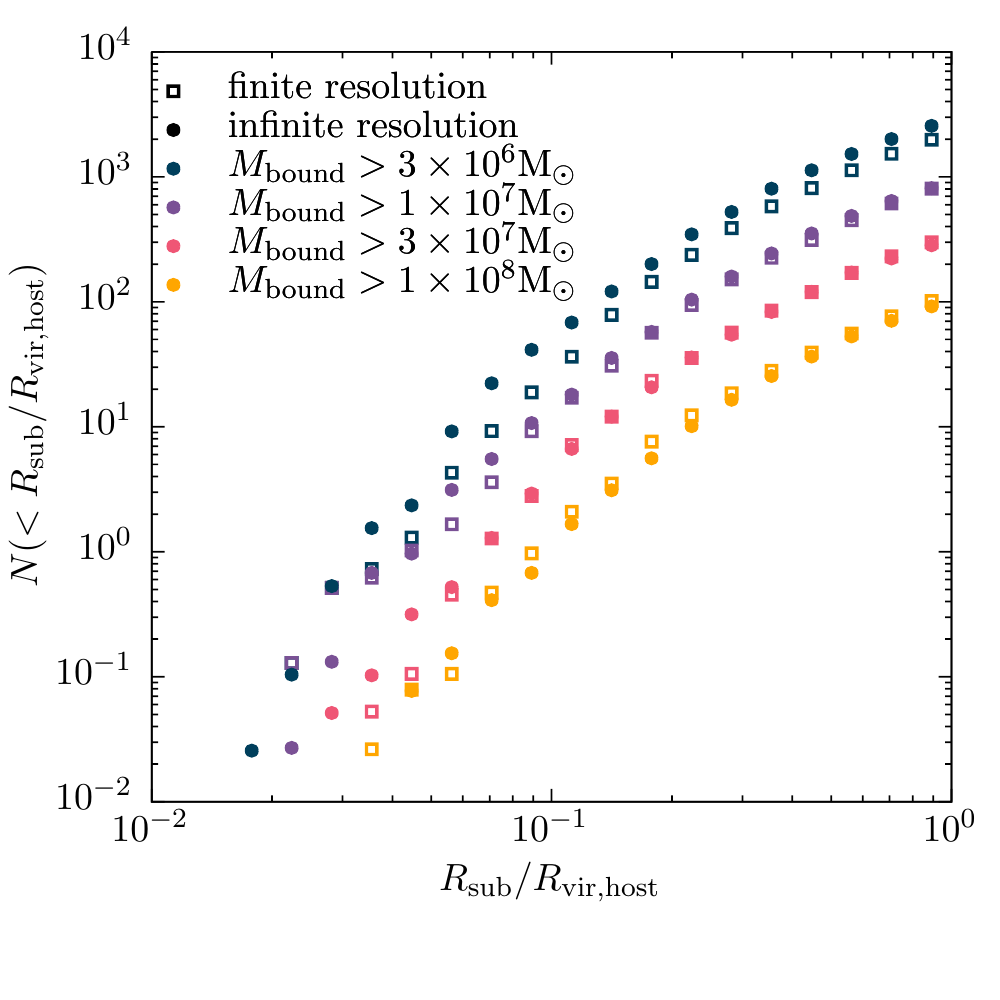}
\end{tabular}
\caption{Radial distributions of subhalos normalized to the virial radius of the host halo. \emph{Left panel:} The distribution for all subhalos more massive than $6.9 \times 10^{-5} M_\mathrm{host}$. Grey points show results from the Caterpillar \protect\citep{2016ApJ...818...10G} LX14 simulations, while blue and green points show results from this work computed with finite resolution (matched to Caterpillar LX14) and infinite resolution respectively. For reference, the dashed black line shows the corresponding NFW distribution \protect\citep{1996ApJ...462..563N} for the mean concentration, $c=10.903$ \protect\citep{2016ApJ...818...10G}, of the Caterpillar halos. \emph{Right panel:} Radial distributions of subhalos from this work computed with infinite (filled circles) and finite (open squares) resolution. Results are shown for various subhalo bound mass thresholds as indicated in the panel.}
\label{fig:subhaloRadialFunction}
\end{figure*}

Figure~\ref{fig:subhaloRadialFunction} shows the resulting radial distribution of subhalos (normalized by the host halo virial radius) for both infinite and finite resolution cases, for four different subhalo bound mass thresholds.

Two clear trends are seen:
\begin{enumerate}
\item The effects of finite resolution are negligible at the highest masses shown ($10^8\mathrm{M}_\odot$, corresponding to around 3,300 particles in Caterpillar LX14), but become much more significant for the lowest mass sample of subhalos shown ($3\times 10^6\mathrm{M}_\odot$, corresponding to around 100 particles).
\item The effect of finite resolution becomes larger at smaller radii, where tidal effects are stronger. For the $3\times 10^6\mathrm{M}_\odot$ sample the reduction in the number of subhalos is about a factor 1.8 at 10\% of the virial radius and a factor 2.8 at 2\% of the virial radius.
\end{enumerate}

\begin{figure}
\includegraphics[width=85mm]{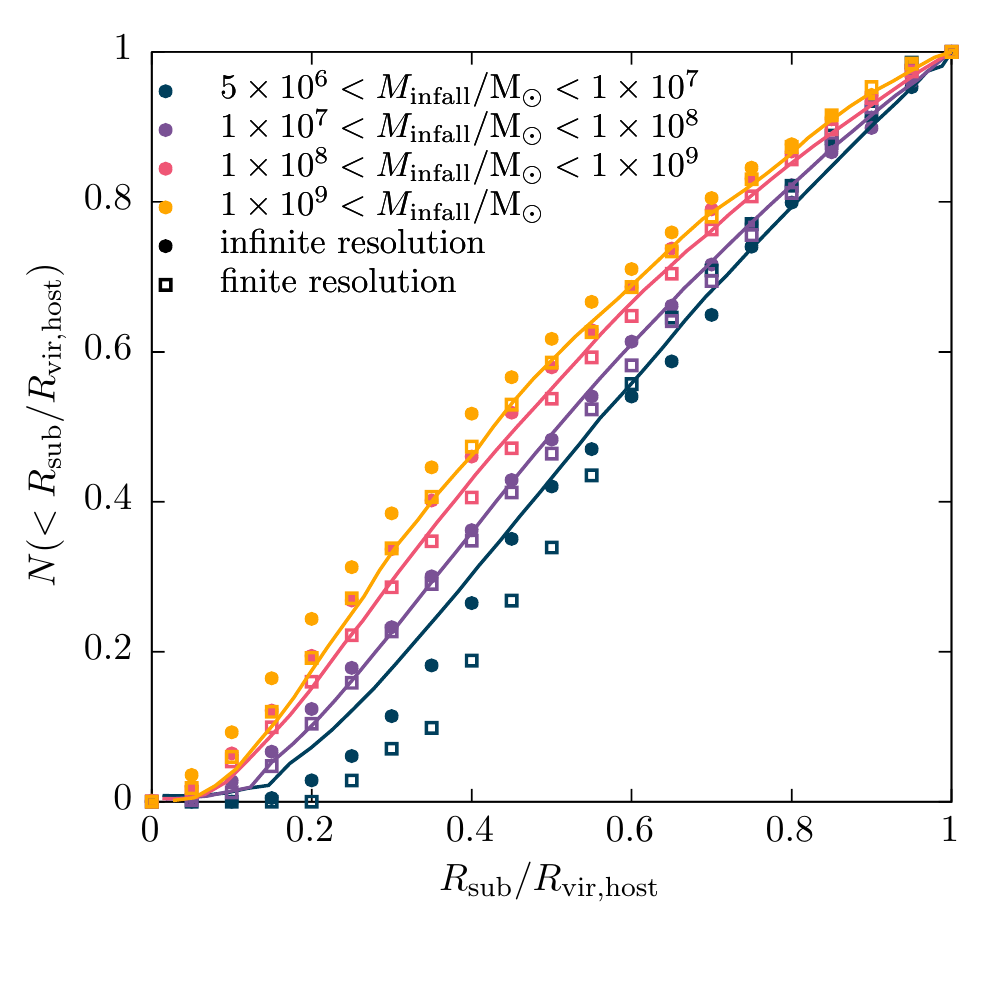}
\caption{Radial distributions of subhalos normalized to the virial radius of the host halo. Results from this work are shown by filled circles for the case of infinite and open squares for the case of finite resolution. Lines show the corresponding results from the Caterpillar LX14 simulations as reported by \protect\cite{2021arXiv211204511M}. Results are shown for various subhalo infall mass thresholds as indicated in the panel.}
\label{fig:subhaloRadialFunctionInfall}
\end{figure}

In figure~\ref{fig:subhaloRadialFunctionInfall} we show subhalo radial distributions selected by peak mass (rather than bound mass) which, in Galacticus, corresponds to the mass of the subhalo at infall. We choose mass intervals matched to those considered by \cite{2021arXiv211204511M} for the same Caterpillar LX14 simulations, and show their results as lines. While the calibration carried out by \cite{2020MNRAS.498.3902Y} did not consider the radial distribution of subhalos, figure~\ref{fig:subhaloRadialFunctionInfall} shows that Galacticus nevertheless produces a reasonable match to these radial distributions as a function of infall mass (except for the lowest mass bin considered in which the Galacticus prediction is less centrally-concentrated than the Caterpillar LX14 results). In particular, the qualitative and quantitative trends with infall mass are reasonably well reproduced. The effects of artificial mass loss and disruption are clearly visible in this plot (becoming more pronounced at smaller radii) even for the more massive subhalos. For example, at 20\% of the virial radius, artificial numerical effects reduce the abundance of subhalos with $M_\mathrm{infall} > 10^9\mathrm{M}_\odot$ by around 25\%.

\section{Discussion}\label{sec:discussion}

We have shown that an improved version of the tidal heating model described by \citeauthor{2001ApJ...559..716T}~(\citeyear{2001ApJ...559..716T}; see also \protect\citealt{1999ApJ...514..109G,2014ApJ...792...24P}), which includes the second-order heating term in the impulse approximation, can accurately match the tidal tracks of subhalos measured from high-resolution N-body simulations. Using this model it is possible to accurately follow the evolution of the density profile of subhalos in semi-analytic models such as those described by \cite{2001ApJ...559..716T}, and \cite{2014ApJ...792...24P}. This is of key importance for many astrophysical observables, including the frequency of gaps in stellar streams  \citep{Carlberg_2012,10.1093/mnras/stw1957}, gravitational lensing by dark substructures \citep{2020MNRAS.492L..12G}, and the rate of dark matter annihilation \citep{2019PhRvD.100f3505D}.

Furthermore, we have demonstrated that this tidal heating model, when applied to halos with a constant density core tuned to match the limited resolution of N-body simulations, can accurately reproduce deviations from the expected tidal track due to finite resolution effects, thereby accurately capturing artificial disruption of subhalos.

Applying this model of artificial disruption to a cosmological system of subhalos we find that results from N-body simulations are likely to be significantly biased for subhalos resolved by fewer than 1,000 particles in the inner regions of their host halo. Figure~\ref{fig:subhaloRadialFunction} suggests that to get accurate subhalo statistics at 1\% of the virial radius at least 1,000 particles per subhalo are required. This is just to get global properties of subhalos (e.g. their masses) correct. Accurately probing the details of internal structure (e.g. $V_\mathrm{max}$) would require even more particles---for example \cite{2021MNRAS.505...18E} note that all of their simulated subhalos show deviations from convergence once reduced to fewer than 3,000 particles.

\cite{2019MNRAS.490.2091G} calibrated models of structural evolution and tidal tracks to the DASH library \citep{2019MNRAS.485..189O}, which achieves sufficiently high resolution to avoid the artificial disruption found in cosmological simulations. This approach, being calibrated directly to the results of high-resolution N-body simulations, is likely more accurate than ours (as it does not have to make simplifying assumptions of spherical symmetry, no shell crossing, etc.), but the approach described here has the advantage that it provides some valuable physical insight into the origin of subhalo tidal tracks, and can be applied to initial density profiles other than NFW.

Using the models calibrated by \cite{2019MNRAS.490.2091G}, \cite{2021MNRAS.503.4075G} study the impact of artificial disruption in the Bolshoi simulations \citep{2011ApJ...740..102K}. They find that artificial disruption leads to an approximately 10\% suppression of the subhalo fraction within the virial radius, and around 20\% suppression of the subhalo mass function. They find the effect is greater at smaller host-centric radii, with the subhalo mass function being reduced by a factor of 2 within 10\% of the host virial radius. This is in excellent agreement with the results of this work (as shown in Figure~\ref{fig:subhaloRadialFunction}). 

We note that the subhalo orbit physics model in Galacticus was calibrated by \cite{2020MNRAS.498.3902Y} to match the results of the Caterpillar \citep{2016ApJ...818...10G} and Elvis \citep{2014MNRAS.438.2578G} simulations. In \cite{2020MNRAS.498.3902Y} the original tidal heating model (containing only the first-order heating term) was used, and subhalos were represented by NFW density profiles, with no constant density core to account for the finite resolution of the simulations to which it was being calibrated.

Presumably there is a unique, infinite resolution model of subhalo orbital physics\footnote{Of course, we do not claim that our current model for subhalo orbital physics captures all of the relevant physics, merely that such a model should exist.} (i.e. the models of tidal mass loss, heating, dynamical friction, etc.), which, when augmented with a treatment of the appropriate finite resolution effects, would accurately match the results of any given cosmological N-body simulation. Of course, the magnitude of those finite resolution effects will be a function of the particle mass, and softening length in each specific N-body simulation.

With the improved tidal heating model, and a treatment of finite resolution effects now implemented in Galacticus, we can therefore repeat the calibration process of \cite{2020MNRAS.498.3902Y} for the subhalo orbit physics (tidal mass loss, tidal heating, and dynamical friction). Once recalibrated with finite resolution effects included, those finite resolution effects can then be ``switched off'', resulting in a model which will approximate subhalo tidal evolution in the limit of infinite resolution. We leave this recalibration of the subhalo orbit physics to a subsequent paper.

\section{Conclusions}\label{sec:conclusions}

The tidal evolution of subhalos---including changes in their density profile and their ultimate survival or destruction---is important for modeling many astrophysical observables, including substructure lensing, the frequency of gaps in stellar streams, and the rate of dark matter annihilation \citep{Carlberg_2012,10.1093/mnras/stw1957,2019PhRvD.100f3505D,2020MNRAS.492L..12G}.

Prior work \citep[e.g.][]{2008ApJ...673..226P,2021MNRAS.505...18E} has shown that subhalos initially described by an NFW density profile follow a well-defined ``tidal track'' in the space of $(r_\mathrm{max}/r_\mathrm{max,0},v_\mathrm{max}/v_\mathrm{max,0})$ as they evolve inside the potential of their host halo. This tidal track is largely independent of the details of the satellite orbit. In this work we show that the simple tidal heating model of \cite{2001ApJ...559..716T} can accurately reproduce these tidal tracks once the second-order heating term is explicitly included.

Importantly, in addition to matching the tidal tracks measured from N-body simulations, this model provides a way to compute the full radial dependence of the density profile in a tidally-heated subhalo. Furthermore, as the model is based upon simple physical arguments \citep{1999ApJ...514..109G} it can be applied to initial density profiles other than NFW---we will explore the consequences of this model for such profiles in a subsequent work.

Future tests of this model could be made using N-body simulations of a more diverse set of density profiles (for example, with different inner and outer slopes, including a central core; \citealt{2010MNRAS.406.1290P}). In future tests of our model for the effects of finite resolution, properties of the initial core could be extracted directly from idealized N-body simulations---allowing more direct validation of our model for $\Delta x$.

In conclusion, the simple tidal heating model presented in this work provides an accurate and rapid means to follow the structural evolution of dark matter subhalos in (semi-)analytic models.

\section*{Acknowledgements}

We acknowledge invaluable conversations with Ethan Nadler, Fangzhou Jiang, Rapha\"el Errani, Jorge Pe\~narrubia, Annika Peter, and Shengqi Yang. 

\section*{Data Availability}

The datasets used in this work were derived from sources in the public domain, specifically the Galacticus semi-analytic model at \href{https://github.com/galacticusorg/galacticus}{\tt https://github.com/galacticusorg/galacticus}. The parameter files used to generate the data underlying this article are available in Zenodo, at \href{https://doi.org/10.5281/zenodo.6612542}{\tt https://doi.org/10.5281/zenodo.6612542}.

\bibliographystyle{mnras}
\bibliography{tidalTracks} 
\bsp	
\label{lastpage}
\end{document}